\documentclass[12pt]{article}
\usepackage{amsmath}
\usepackage{amssymb,amsfonts}
\usepackage[all]{xy}
\usepackage{graphicx}
\usepackage[parfill]{parskip}
\usepackage[usenames]{color}
\usepackage{tikz}
\usetikzlibrary{shapes.geometric}
\usetikzlibrary{snakes}
\usepgflibrary{arrows}

\numberwithin{equation}{section}
\numberwithin{table}{section}

\def\inttorus{\int\displaylimits_{\begin{tikzpicture}
\node [draw, trapezium, minimum width=0.5cm, trapezium left angle=60, trapezium right angle=120] {} ;
\end{tikzpicture}}}
\def\btr{\int\displaylimits_{\resizebox{0.6cm}{!}{\begin{tikzpicture}[>=triangle 60]
\node(a) {};
\node(b) at (2,2) {};
\draw[snake=zigzag] (a) -- (b);
\draw [->, very thick] (0,0) arc (270:360:2.2cm);
\end{tikzpicture}}}}

\def\atl{\int\displaylimits_{\resizebox{0.6cm}{!}{\begin{tikzpicture}[>=triangle 60]
\node(a) {};
\node(b) at (2,2) {};
\draw[snake=zigzag] (a) -- (b);
\draw [<-, very thick] (-.2,-.2) arc (180:90:2.2cm);
\end{tikzpicture}}}}

\def\a{\alpha}

\newcommand{\be}{\begin{equation}}
\newcommand{\ee}{\end{equation}}

\newcommand{\Li}{\mathrm{Li}}
\newcommand{\re}{\mathrm{Re}}
\newcommand{\im}{\mathrm{Im}}

\let\oldsqrt\sqrt
\def\sqrt{\mathpalette\DHLhksqrt}
\def\DHLhksqrt#1#2{%
\setbox0=\hbox{$#1\oldsqrt{#2\,}$}\dimen0=\ht0
\advance\dimen0-0.2\ht0
\setbox2=\hbox{\vrule height\ht0 depth -\dimen0}%
{\box0\lower0.4pt\box2}}

\def\beq{\begin{eqnarray}}
\def\eeq{\end{eqnarray}}
\def\ba{\begin{eqnarray}}
\def\ea{\end{eqnarray}}
\def\a{\alpha}

\def\ep1{\epsilon_1}
\def\eps2{\epsilon_2}

\newcommand{\IZ}{\mathbb{Z}}
\newcommand{\IC}{\mathbb{C}}

\newcommand{\IR}{\mathbb{R}}

\newcommand{\res}{\mathrm{Res\,}}

\newcommand{\nn}{\nonumber}

\newcommand{\cW}{{\cal W}}
\newcommand{\cN}{{\cal N}}

\newcommand{\cF}{{\cal F}}

\newcommand{\tphi}{\tilde{\phi}}
\newcommand{\tw}{\tilde{w}}
\newcommand{\tz}{\tilde{z}}

\newcommand{\Vt}{V}

\begin{document}

\thispagestyle{empty}

\vskip 3cm
\noindent
\begin{center}
{\LARGE \bf Quantum geometry from the toroidal block}
\end{center}
\vskip .4cm
\begin{center}
\linethickness{.06cm}
\line(1,0){447}
\end{center}
\vskip .5cm
\noindent
{\large \bf Amir-Kian Kashani-Poor and Jan Troost}
 
\noindent

\vskip 0.15cm
{\em \hskip -.05cm Laboratoire de Physique Th\'eorique\footnote{Unit\'e Mixte du CNRS et
    de l'Ecole Normale Sup\'erieure associ\'ee \`a l'Universit\'e Pierre et
    Marie Curie 6, UMR
    8549.}}
    \vskip -0.22cm

{\em \hskip -.05cm Ecole Normale Sup\'erieure}
   \vskip -0.22cm

{\em \hskip -.05cm 24 rue Lhomond, 75005 Paris, France}

\vskip 1cm

\vskip0cm

\noindent {\sc Abstract:} 
We continue our study of the semi-classical (large central charge)
expansion of the toroidal one-point conformal block in the context of
the 2d/4d correspondence. We demonstrate that the Seiberg-Witten curve
and ($\epsilon_1$-deformed) differential emerge naturally in conformal
field theory when computing the block via null vector decoupling
equations. This framework permits us to derive
$\epsilon_1$-deformations of the conventional relations governing the
prepotential. These enable us to complete the proof of the quasi-modularity of the coefficients of the conformal block in an expansion around large exchanged conformal dimension. We furthermore derive these relations from the semi-classics of exact conformal field theory quantities, such as braiding matrices and the S-move kernel. In the course of our study, we present a new proof of Matone's relation for $\cN=2^*$ theory.

\vskip 1cm

\pagebreak

\tableofcontents

\section{Introduction}
The semi-classical limit of conformal blocks has taken on new
significance with the advent of the 2d/4d correspondence
\cite{Alday:2009aq} between two-dimensional conformal field theory and
four-dimensional $\cN=2$ gauge theory. Holomorphic amplitudes
$F^{(n,g)}$ specifying the dynamics of the gauge theory, such as the
prepotential $F^{(0,0)}$, arise as coefficients in an asymptotic
expansion of the blocks in this limit. Several methods of computation
precede the 2d/4d correspondence: among these, the generalized holomorphic anomaly equations
compute the $F^{(n,g)}$ directly \cite{Bershadsky:1993cx,Krefl:2010fm,Huang:2010kf,Huang:2011qx}, while localization calculations
yield their generating function, the Nekrasov partition function \cite{Nekrasov:2002qd}. The conformal field theory perspective opens up a new avenue for the
computation of the amplitudes $F^{(n,g)}$ based on null vector
decoupling equations. In previous work
\cite{KashaniPoor:2012wb,Kashani-Poor:2013oza}, we demonstrated how to
compute the $F^{(n,0)}$ in this setup, for $\cN=2^*$ and $N_f=4$ gauge
theory. Here, for the case of $\cN=2^*$, we will demonstrate how to
rederive the Seiberg-Witten approach \cite{Seiberg:1994rs,Seiberg:1994aj} for computing the prepotential of
the gauge theory within this framework, and extend these techniques to
the generating function $F= \sum F^{(n,0)} \epsilon_1^{2n}$. In
particular, an $\epsilon_1$-deformed Seiberg-Witten differential will
emerge naturally in this setup. As Seiberg and Witten geometrize the
problem of computing the prepotential by relating it to an elliptic
fibration over moduli space, and as deformation by $\epsilon_1$ can be
interpreted as a quantum deformation from an integrable systems
perspective \cite{Nekrasov:2009rc}, such an extension can be said to describe quantum
geometry. In the course of our investigations, we will prove the
quasi-modularity of the $F^{(n,0)}$, a property we had observed
experimentally in \cite{KashaniPoor:2012wb}. Our argument includes the
first proof of Matone's relation for a superconformal theory in the
Seiberg-Witten framework. A proof using localization methods has
appeared in \cite{Flume:2004rp}.

A second theme of this work is extracting results we obtain from the
semi-classical solution of null vector decoupling equations directly
from the semi-classics of known exact relations in conformal field
theory. We thus study the dual period of the $\epsilon_1$-deformed
Seiberg-Witten differential by using braiding
matrices to compute the monodromy behavior of the toroidal block, and also extract the transformation properties of the generating function
$F$ under S-duality from the integral kernel implementing the S-move
on the block.

Our interest in studying the amplitudes $F^{(n,g)}$ in the context of
exact conformal field theory results stems from their interpretation
as limits of topological string amplitudes. While the topological
string partition function from a worldsheet perspective is merely a
generating function for these amplitudes, it acquires a
non-perturbative definition in the light of the 2d/4d
correspondence. Our hope is therefore that a careful study of how
semi-classical results emerge from exact quantities in conformal field
theory will help clarify non-perturbative aspects of topological
string theory.

Our paper is organized as follows. In section \ref{toroidalblock}, we
review the toroidal one-point block of conformal field theory. In
section \ref{nullvectordecoupling}, we revisit the construction of the
semi-classical conformal block through a WKB solution of the null
vector decoupling equation. We show how the Seiberg-Witten curve and ($\epsilon_1$-deformed) differential arise naturally in this framework, allowing us to demonstrate that up to
exceptional leading terms, the expansion of the logarithm of the
block is in terms of quasi-modular forms. We offer a proof of Matone's relation for $\cN=2^*$ theory in the course of the argument.
The exact formulae for the braiding matrices and the
modular S-move on the toroidal one-point block are reviewed in section
\ref{smove}, and used to rederive some of the results of section \ref{nullvectordecoupling} upon saddle point approximation. We conclude in section
\ref{conclusions}.

\section{The toroidal one-point block}
\label{toroidalblock}
In this section, we will briefly review the definition of the one-point conformal block of
two-dimensional conformal field theory, and of the two-point conformal
block that includes one degenerate insertion. These map to the instanton partition function of $\epsilon$-deformed $\cN=2^*$ gauge theory \cite{Alday:2009aq}, the latter in the presence of a surface operator \cite{Alday:2009fs}.

\subsection{The one-point function and conformal block}
In a two-dimensional conformal field theory, the expectation value of
an operator $V_{h_m}$ with conformal dimension $h_m$ inserted on a
torus with complex structure parameter $\tau$ can be decomposed in terms
of the three-point functions $C_{h_m,h}^h$ of the theory and the toroidal
one-point blocks ${\cal F}^h_{h_m}$,
\begin{eqnarray}  \label{one_point}
  \langle \Vt_{{h}_m} \rangle_\tau &=&
\sum_{{h}} C^{h}_{{{h}_m} ,{h}} (q \bar{q})^{{h}-\frac{c}{24}}
|{\cal F}_{{h}_m} ^{h} (q)|^2 \,.
\end{eqnarray}
The sum here is over all the primary fields, of conformal
dimensions $h$, in the spectrum of the theory.
In terms of chiral vertex operators, the one-point toroidal conformal block can be represented as
the trace
\ba
 q^{h- \frac{c}{24}}\cF_{h_m}^h  &=& \mathrm{Tr}_h \, \Big( q^{L_0-\frac{c}{24}} \hspace{-0.6cm}
\begin{aligned}
\xymatrix@!0@M0pc@W0pc@H0pc{  & &   & & &\\
      &    \ar@{-}'[rrrr]^>(.9){h}^<(.1){h}
& &     \ar@{-}[u]^{h_m} & 
& \Big)
\\
& & & & & 
}
\end{aligned} \\
&=&
\begin{aligned}
\xymatrix@!0@M0pc@W0pc@H0pc{  & &  & & & \\
       &   & \ar@{-}`[rrr]
`[rrrd] `[ld]^{h}  `[l]  []
&  \ar@{-}[u]^{h_m} &   &
\\
& & & & & 
}
\end{aligned}  
\ea
The trace is taken over the primary state $| h \rangle$ and all of its Virasoro descendants.

\subsection{The two-point  block including one degenerate insertion}
To study the one-point conformal block, we will take advantage of the
null vector decoupling equation satisfied by the two-point toroidal
block with the additional insertion chosen to be degenerate at level
two, with weight denoted $h_{(2,1)}$. In addition to the dependence on the two external weights $h_m$ and
$h_{(2,1)}$, this block requires specifying two internal momenta $h$
and $h_{\pm}$, in accord with the diagram
\be \label{2pt_deg}
\cF_{h_m, h_{(2,1)}}^{h, h_\pm} =
\begin{aligned}
\xymatrix@!0@M0pc@W0pc@H0pc{  & &  & & & \\
       &   & \ar@{-}`[rrr]  ^{h_\pm} 
`[rrrd] `[ld]^{h}  `[l]  []   \ar@{-}[u]^{h_m}
&  & \ar@{-}[u]_{h_{(2,1)}}   &
\\
& & & & &
}
\end{aligned} 
\ee 
Note that due to the degenerate nature of the primary with weight $h_{(2,1)}$, 
this block is non-vanishing only for two choices of internal weight $h_{\pm}$ as a function of the
exchanged conformal weight $h$.

\subsection{The variable map and limits}

Our calculations will take place purely within conformal field
theory. By the 2d/4d correspondence \cite{Alday:2009aq}, the one-point
toroidal block thus obtained is equal to the instanton partition
function of $\cN=2^*$ gauge theory, upon the following identification of
variables:
\begin{eqnarray}
c &=& 1 + 6 Q^2 ,  \qquad Q = b+ b^{-1} , \qquad
b = \sqrt{\frac{\epsilon_2}{\epsilon_1}}  \,,
\nonumber \\
{h}_m &=& \frac{Q^2}{4} - \frac{m^2}{\epsilon_1 \epsilon_2} \, ,  \qquad
{h} = \frac{Q^2}{4} - \frac{a^2}{\epsilon_1 \epsilon_2} \,. \label{id_par}
\end{eqnarray}
The central charge $c$ of the theory is parameterized by the
numbers $Q$ or $b$. In gauge theory, $a$ and $m$ are the vacuum expectation value of the adjoint vector
multiplet scalar and the mass
parameter of the adjoint matter multiplet respectively. In the refined topological string context \cite{Iqbal:2007ii}, the $\epsilon_i$ deformation parameters are related to the string coupling $g_s^2 = \epsilon_1 \epsilon_2$ and the expansion
parameter $s =
(\epsilon_1 + \epsilon_2)^2$.

We will be working in the semi-classical limit of large central charge $c \rightarrow \infty$ ($b \rightarrow 0$) and conformal dimensions, with $h/c \rightarrow \infty$ and $h/h_m \rightarrow \infty$. In terms of gauge theory parameters, this corresponds to the small $\epsilon_i$, large vacuum expectation
value limit $\epsilon_2/\epsilon_1 \rightarrow 0$, $\epsilon_1/a \ll 1$, and $m/a  \ll 1$.

\section{Seiberg-Witten geometry and S-duality from null vector decoupling}
\label{nullvectordecoupling}
In this section, we analyze the properties of the semi-classical solutions to the 
second order null vector decoupling equation satisfied by the two-point block.  In particular, we identify the
Seiberg-Witten data that determine the solution within conformal field theory. This permits us to prove the quasi-modularity of the expansion coefficients of the semi-classical solution,
observed experimentally in \cite{KashaniPoor:2012wb}, to all orders. In the process, we provide a proof of the Matone relation for $\cN=2^*$ theory.

\subsection{The null vector decoupling equation}
Our principal strategy for computing the one-point toroidal block and
determining its transformation properties is, as in \cite{Fateev:2007qn,Marshakov:2010fx,KashaniPoor:2012wb,Kashani-Poor:2013oza}, to consider an
additional, second order degenerate insertion on the torus. Due to the degeneracy
of the insertion, the resulting two-point function satisfies a differential equation, the null vector decoupling
equation, which upon rescaling of the two-point function,
\be 
\label{rescaling}
\Psi(z |\tau) =  \theta_1(z|\tau) ^{-\frac{b^2}{2}} \eta(\tau)^{-2({h}_m - b^2 -1)} Z \langle \Vt_{h_{(2,1)}}(z) \Vt_{{h}_m} (0) \rangle_\tau \,,
\ee
takes the simple form \cite{Marshakov:2010fx,KashaniPoor:2012wb}
\be \label{diff_equ_psi}
\left[-\frac{1}{b^2} \partial_z^2 - \big(\frac{1}{4b^2} -
  \frac{m^2}{\epsilon_1 \epsilon_2} \big) \wp(z) \right] \Psi(z|\tau) = 2 \pi
i \partial_\tau \Psi(z|\tau) \,.
\ee 
The function $\wp$ is the Weierstrass $\wp$-function associated to the torus of periods $(1,\tau)$ on which the conformal field theory lives, and the function $Z$ is the partition function.
Imposing the monodromy \cite{Mathur:1988yx,KashaniPoor:2012wb}
\be \label{bc}
\Psi(z+1) = e^{\pm 2 \pi i \frac{a}{\epsilon_1}}\Psi(z)
\ee
on the solution to the differential equation permits us to project
onto the conformal block $\cF_{h_m,h_{(2,1)}}^{h,h_\pm}$, in the
notation of (\ref{2pt_deg}). The analysis up to this point is
exact. To extract the one-point conformal block of interest from the
rescaled two-point function $\Psi$, we need to take the semi-classical $\epsilon_2/\epsilon_1
\rightarrow 0$ limit, in which the degenerate insertion becomes light
and its contribution to the conformal block is multiplicative (as can be seen in a
semi-classical analysis of Liouville theory (see e.g. \cite{Harlow:2011ny})).
This motivates the factorized
ansatz
\begin{eqnarray} \label{fact_ansatz}
\Psi(z|\tau) &=& \exp \left[ \frac{1}{\epsilon_1 \epsilon_2} 
{\cal F}(\tau) + \frac{1}{\epsilon_1} {\cal W}(z|\tau)  \right] 
\end{eqnarray}
with functions $\cF$ and $\cW$ that are independent of $\epsilon_2$. In terms of this ansatz, the
one-point block is given by
\be \label{one_point_F}
\lim_{\epsilon_2 \rightarrow 0} Z\,\langle V_{h_m} \rangle \Big\vert_{h} = \exp\left[\frac{1}{\epsilon_1 \epsilon_2} \left( \cF +2( -m^2 + \frac{\epsilon_1^2}{4} ) \log \eta \right) \right] \,.
\ee
The null vector decoupling equation (\ref{diff_equ_psi}) evaluated on the ansatz (\ref{fact_ansatz}) yields the equation
\be
-\frac{1}{\epsilon_1} \cW''(z|\tau) - \frac{1}{\epsilon_1^2} \cW'(z|\tau)^2 +  \left( \frac{1}{\epsilon_1^2} m^2 -\frac{1}{4} \right) \wp(z)  =(2 \pi i)^2 \frac{1}{\epsilon_1^2} q \partial_q {\cal F} (\tau)+  \frac{\epsilon_2}{\epsilon_1^2} 2 \pi i \partial_\tau \cW(z|\tau)  \,,   \label{diff_equ_fact}
\ee
while the boundary condition (\ref{bc}) projecting onto the desired conformal block maps to the condition
\be   \label{bc_fact}
\cW(z+1) - \cW(z) = \pm 2 \pi i a \,.
\ee

In \cite{KashaniPoor:2012wb}, we solved this differential equation (dropping the linear term in $\epsilon_2$) in a formal $\epsilon_1$-expansion of $\cF$ and $\cW$, 
\be
 {\cal F}(\tau) =
\sum_{n=0}^\infty {\cal F}_n(\tau) \epsilon_1^n \,, 
\quad {\cal  W}(z|\tau) = \sum_{n=0}^\infty {\cal W}_n(z|\tau) \epsilon_1^n \,,  
\ee 
and demonstrated to a given order that the coefficients $\cF_n$ of the
non-convergent expansion reproduce the modular results obtained from
the holomorphic anomaly equations \cite{Huang:2011qx} and localization calculations \cite{Billo:2013fi},
\be
F^{(n,0)} = \cF_{2n} \,.
\ee

\subsection{The amplitudes from generalized period integrals}

\subsubsection{The generalized Seiberg-Witten differential}
Combining equations (\ref{diff_equ_fact}) and (\ref{bc_fact}) yields the equation
\be   \label{genper}
\int_0^1 \sqrt{m^2 \wp - (2 \pi i)^2 q\partial_q \cF - \epsilon_1 \cW'' -  \epsilon_1^2\, \frac{\wp}{4} } \,dz= \pm 2 \pi i a  \,.
\ee
Given that the variable $a$ maps to the vacuum expectation value of the adjoint scalar field in $\cN=2^*$ gauge theory, we wish
to interpret 
\be  \label{def_def_diff}
\lambda := \sqrt{m^2 \wp - (2 \pi i)^2 q\partial_q \cF - \epsilon_1 \cW'' -\epsilon_1^2\, \frac{\wp}{4} } \, dz
\ee
as a generalized $\epsilon_1$-dependent Seiberg-Witten
differential. We will justify this interpretation by
demonstrating that the $B$-cycle period of $\lambda$ computes the
$a$-derivative of the generalized prepotential $\cF$,
\be
2 \pi i \,a_D := \oint_B \lambda = -\frac{1}{2}\frac{\partial \cF}{\partial a}  \,.
\ee
This equation is to be interpreted as an equality of formal power series in $\epsilon_1$.

\subsubsection{Seiberg-Witten data and  proof of the 
Matone relation}
To leading order in $\epsilon_1$, we find
\be \label{lambda0}
\lambda_0 := \sqrt{ m^2 \wp - u} \,dz \,, \quad u := 2\pi i \, \partial_\tau \cF_0  \,.
\ee
We first determine the Riemann surface on which the square root appearing in the differential
is single-valued. The Weierstrass $\wp$-function provides a two-to-one mapping from the torus
to the sphere. The equation $m^2 \wp=u$ hence has two solutions on the
torus. Single-valuedness of the differential $\lambda_0$ requires a branchcut
connecting them. The natural home of the differential is therefore a curve
of genus two. The curve degenerates at $ \frac{u}{m^2} = e_i$, with
$e_i$ any of the half-periods of the domain torus of $\wp$. By writing
\be
t^2 = m^2 \wp - u \,,  \quad y^2 = 4 \prod_{i=1}^3 ( \wp - e_i)  \,,
\ee
we can present the genus two Riemann surface in its hyperelliptic form,
\be  \label{gen_2_sw}
y^2 = 4\prod_i \left( \frac{t^2 + u}{m^2}  - e_i \right) 
\ee
with holomorphic one-forms $\omega_i$
\be
\omega_1 = 2 \,\frac{dt}{y} = \frac{m^2 dz}{\sqrt{m^2 \wp - u}} \,, \quad   \omega_2 = 2 \,\frac{t dt}{y} = m^2 dz \,.
\ee
We will denote the cycles on the genus two surface as $A_{\pm}$ and $B_{\pm}$, with $A$, $B$ specifying the cycle on each sheet (which has the topology of a torus) and $\pm$ specifying the sheet. Thus,
\be \label{periods_sign}
\oint_{A_+, B_+} \omega_i = -\oint_{A_-, B_-} \omega_i  \,,
\ee
and in particular,
\be \label{sw_pm}
\oint_{A_\pm} \lambda_0 = \pm 2 \pi i a  \,.
\ee
Defining the dual period
\be
2 \pi i \, a_D^0 := \oint_{B_+} \lambda_0  \,,
\ee
we will prove that
\be  \label{mat_tau}
2 \pi i \frac{\partial a_D^0}{\partial \tau} =- \frac{1}{4 \pi i}\frac{\partial u}{\partial a}  \,,
\ee
where the $a$-dependence of $u$ is determined by equation
(\ref{sw_pm}). Integrating this relation with regard to $\tau$, we
will thus obtain the equality of the $B_+$ period of the logarithm of
the semi-classical two-point conformal block and the $a$-derivative of
the logarithm of the one-point block, up to a $\tau$ independent
function. 

\subsubsection*{The gauge theoretic point of view}
From a gauge theory perspective, with the modulus $u$ given as in
(\ref{lambda0}) and $\cF_0$ identified as the prepotential of the
gauge theory via the 2d/4d correspondence, this equality demonstrates
that the $a$-derivative of the prepotential is the $B_+$ period of
$\lambda_0$, thus justifying identifying $\lambda_0$ as the
Seiberg-Witten differential on the Seiberg-Witten curve
(\ref{gen_2_sw}).

Note that in the original formalism of Seiberg and Witten, the curve of a rank one gauge
theory has genus one. In \cite{Donagi:1995cf}, precisely the genus two
curve (\ref{gen_2_sw}) appears, with a prescription for recovering the
Seiberg-Witten data from the higher dimensional Jacobian. The interpretation of the full Jacobian is as follows: the ratio
of $B_+$ to $A_+$ period of $\omega_2$ yields the ultraviolet coupling $\tau$
of the theory, while the ratio of the corresponding periods of
$\omega_1$ yields the infrared coupling as determined by $\lambda_0$ as
Seiberg-Witten differential. Exchanging $+$ for $-$ cycles merely
changes signs in accord with (\ref{periods_sign}). In \cite{Witten:1997sc,Gaiotto:2009we},
the Seiberg-Witten curves of $SU(2)$ superconformal Seiberg-Witten theories
are proposed to generally arise as the double cover of curves
parametrized by the ultraviolet couplings of the theory.

The proof of equation (\ref{mat_tau}) can also be
interpreted from the conventional angle, in which $u$ is a coordinate
on the gauge theory moduli space, a priori unrelated to the
prepotential ${\cal F}_0$. The latter is introduced via its relation to the
dual period of the Seiberg-Witten differential, $\partial {\cal F}_0
/ \partial a = -4 \pi i \,a_D^0$. The equation (\ref{mat_tau}) then becomes the
$a$-derivative of the Matone relation for $\cN=2^*$ gauge theory (up to
an $a$-independent term in ${\cal F}_0$, which carries no physical
interpretation), 
\be \label{matone} 2 \pi i \frac{\partial
  {\cal F}_0(a,\tau)}{\partial \tau} = u\,.  
\ee 
The proof that will follow
hence also provides the first demonstration of this
equation for $\cN=2^*$ purely within the Seiberg-Witten framework. A
demonstration using instanton calculus has appeared in
\cite{Flume:2004rp}.

\subsubsection*{The proof}

For simplicity of notation, we will set $m^2=1$ in the following. The $m$ dependence can easily be restored via dimensional analysis, by assigning mass dimension 2 to $u$.

The proof we present is a variant of the proof of the Riemann bilinear identity. We define the function
\be
\eta_0(z) = \int^z \frac{1}{\sqrt{\wp - u}} \,dz'  \,.
\ee
We will calculate the integral of
\be  \label{kernel}
\eta_0(z) \partial_\tau  \lambda_0
\ee
along the parallelogram in the complex plane spanned by $1$ and $\tau$ in two ways: by 
integrating along the edges of the parallelogram, and alternatively, by contracting the contour
inside the torus to hug the branch cut.
For the first method, we will make use of the identities (see e.g. \cite{Zabrodin:2011fk})
\be \label{dtaup}
\partial_\tau \wp(z+1,\tau) = \partial_\tau \wp(z,\tau)  \,, \quad  \partial_\tau \wp(z+\tau,\tau) = \partial_\tau \wp(z,\tau) - \wp'(z) \,.
\ee 
We will denote the periods of the one-form $\omega_1$ along the cycles $A_+$ and $B_+$
as $\Pi_A$, $\Pi_B$ respectively.
 The evaluation on the parallelogram then proceeds as
\ba
2 \inttorus \eta_0(z) \partial_\tau \lambda_0 &=& \int_0^\tau \frac{\left(\eta_0 \partial_\tau (\wp -u)\right)(z+1) -\left(\eta_0 \partial_\tau (\wp -u)\right)(z)}{\sqrt{\wp - u}} \,dz \nn\\
&&+\int_0^1 \frac{\left(\eta_0 \partial_\tau (\wp -u)\right)(z) -\left(\eta_0 \partial_\tau (\wp -u)\right)(z+\tau)}{\sqrt{\wp - u}} \, dz\nn\\
&=&\int_0^\tau \frac{\left(\eta_0(z+1)-\eta_0(z)\right) \partial_\tau (\wp -u)(z)}{\sqrt{\wp-u}} \,dz \nn\\
&&+\int_0^1 \frac{\eta_0(z) \partial_\tau (\wp -u)(z) -(\eta_0(z)+\Pi_B)( \partial_\tau (\wp -u)(z) - \wp'(z))}{\sqrt{\wp - u}} \,dz \nn\\
&=& \Pi_A \int_0^\tau \frac{\partial_\tau (\wp -u)(z)}{\sqrt{\wp-u}} \,dz + \int_0^1 \frac{(\eta_0 \wp')(z)}{\sqrt{\wp - u}} \,dz \nn\\
&&- \Pi_B \int_0^1 \frac{\partial_\tau (\wp -u)(z)}{\sqrt{\wp-u}} \,dz + \Pi_B \int_0^1 \frac{\wp'(z)}{\sqrt{\wp - u}} \,dz   \,. \label{RBIleading}
\ea
The last two terms in equation (\ref{RBIleading}) vanish. The first of these does so
because $a$ is assumed to be $\tau$ independent. The second term in equation (\ref{RBIleading}) can be further manipulated:
\ba
\int_0^1 \frac{\eta_0 \wp'}{\sqrt{\wp - u}} \,dz &=& 2 \int_0^1 \eta_0 \frac{\partial}{\partial z} \sqrt{\wp - u} \, dz \nn \\
&=& 2 \int_0^1 \frac{\partial}{\partial z} \left( \eta_0 \sqrt{\wp -u} \right) \, dz - 2 \int_0^1 \sqrt{\wp - u} \frac{\partial}{\partial z} \eta_0 \,dz \nn \\
&=& 2 \eta_0 \sqrt{\wp - u} |_0^1 -2 \nn \\
&=& 2 \sqrt{\wp -u} (0) (\eta_0(1)-\eta_0(0)) -2 \nn\\
&=& 2 \Pi_A \sqrt{\wp-u}(0)  - 2  \,.
\ea
We hence find
\be
2 \inttorus \eta_0(z) \partial_\tau \lambda_0  =  2 \Pi_A \partial_\tau \int_0^\tau \sqrt{\wp -u} \, dz -2
\, ,
\ee
which contains the term we wish to evaluate.

Now, we compute the integral of the form (\ref{kernel}) over the
parallelgram again, this time by first contracting the integration
contour to hug the branchcut between the two $\wp$-preimages of $u$:
\ba
2 \inttorus \eta_0(z) \partial_\tau \lambda_0 \,dz
&=& \btr \frac{\eta_0(z_-) \partial_\tau (\wp(z_-) - u)}{\sqrt{\wp(z_-) - u}} dz_- + \atl \frac{\eta_0(z_+) \partial_\tau (\wp(z_+) - u)}{\sqrt{\wp(z_+) - u}} \, dz_+  \nn \\
&=& \btr \frac{ \left( \eta_0(z_-) + \eta_0(z_+) \right) \partial_\tau (\wp(z_-) - u)}{\sqrt{\wp(z_-) - u}} dz_- \nn\\
&=& 2 \eta_0(\wp^{-1}(u)_1) \btr \frac{\partial_\tau (\wp(z_-) - u)}{\sqrt{\wp(z_-) - u}} dz_- \nn\\
&=& \eta_0(\wp^{-1}(u)_1) \inttorus \frac{\partial_\tau (\wp(z) - u)}{\sqrt{\wp(z) - u}} dz \nn\\
&=&  \eta_0(\wp^{-1}(u)_1) \int_0^1 \frac{\wp'}{\sqrt{\wp-u}} \, dz \nn\\
&=&  \eta_0(\wp^{-1}(u)_1) 2 \sqrt{\wp - u} |_0^1 \nn \\
&=& 0 \,,  \ea 
where $\wp^{-1}(u)_1$ denotes the $\wp$-preimage of $u$ at the lower end of the branch cut (with regard to the diagram). We have thus arrived at the equality 
\be \Pi_A
\, \partial_\tau \oint_{B_+} \lambda_0 = 1 \quad \Leftrightarrow \quad
2 \pi i \partial_\tau a_D = -\frac{1}{2} \frac{\partial \partial_\tau
  \cF_0}{\partial a} \quad \Leftrightarrow \quad 2 \pi i a_D =
-\frac{1}{2} \frac{\partial \cF_0}{\partial a} + g(a) \,, \ee
 with
$g(a)$ independent of $\tau$. Note that the differential equation
(\ref{diff_equ_fact}) and its boundary condition (\ref{bc_fact})
determine $\cF$ only up to a $\tau$ independent piece. We are hence
free to define $\cF$ and in particular $\cF_0$ such that $g(a) = 0$,
and hence 
\be \frac{\partial \cF_0}{\partial a} = -2 \oint_{B_+}
\lambda_0 \,.  \label{legendreequal} \ee 
We will return to this point below after extending
this equality beyond leading order in $\epsilon_1$, and determine the necessary
integration constant that must be included in $\cF$ for the equality
(\ref{legendreequal}) to hold.

Strictly speaking, to avoid integrating over the pole of the function $\wp$, we
should shift all contours in our proof by a constant amount. This will eliminate
the  infinities otherwise present in intermediate steps in the calculation.

\subsubsection{Beyond leading order: quantum geometry}  \label{mat_bey_leading}
Our results from \cite{KashaniPoor:2012wb} yield $\cW''$ as a formal power series
\be  \label{form_W}
\frac{1}{a} \cW'' \in   \IC[E_2,E_4,E_6, \wp, \wp'][[\frac{m}{a}]][[\frac{\epsilon_1}{a}]]\,.
\ee
The coefficients of the
formal series in $\frac{\epsilon_1}{a}$ are convergent power series in
$\frac{m}{a}$. Interpreted as equalities between formal power series
in $\frac{\epsilon_1}{a}$,
the above calculation goes through almost unchanged, with the differential $\lambda_0$
replaced by the differential $\lambda$, the function $\eta_0$ replaced by
\be
\eta(z) = \int^z \frac{dz}{\sqrt{(1- \frac{\epsilon_1^2}{4}) \wp - U - \epsilon_1 \cW''}} \,,  \quad U= 2 \pi i \partial_\tau \cF \,,
\ee
and the periods $\Pi_A$, $\Pi_B$ defined as the $A_+$ and $B_+$ periods of
\be
\Omega_1 = \frac{dz}{\sqrt{(1- \frac{\epsilon_1^2}{4}) \wp - U - \epsilon_1 \cW''}}  \,.
\ee
The argument now relies, in addition to the quasi-periodicity properties
(\ref{dtaup}) of the derivative of the Weierstrass function $\partial_\tau \wp$, on the equality
\ba
\partial_\tau \cW''(z+\tau) &=& \partial^E_\tau \cW''(z+\tau) + \partial_\wp \cW''(z+\tau) \partial_\tau \wp(z+\tau) + \partial_{\wp'} \cW''(z+\tau) \partial_\tau \wp'(z+\tau) \nn\\
&=& \partial_\tau \cW''(z) - \partial_\wp \cW''(z) \wp'(z) - \partial_{\wp'} \cW''(z) \wp''(z) \nn \\
&=& \partial_\tau \cW''(z) - \cW'''(z) \,. 
\ea
The notation $\partial^E_\tau \cW''$ is used to indicate the
derivative of $\cW''$ with regard to the $\tau$-dependence of the
quasi-modular forms in the presentation (\ref{form_W}) of $\cW''$. We
can thus establish the equality between the $a$-derivative of $\cF$
and a period integral over a formal power series involving the $\tau$-derivative of $\cF$, \be \label{b_per} \frac{\partial \cF}{\partial a}
= - 2 \oint_{B_+} \sqrt{ \wp - 2 \pi i \partial_\tau \cF - \epsilon_1
  \cW'' -\epsilon_1^2\, \frac{\wp}{4} } \,dz + G(a) \,, \ee with
$G(a)$ a function independent of $\tau$. As above, we wish to define
$\cF$ to incorporate $G(a)$. To specify the ensuing integration
constant in passing from $\partial_\tau \cF$ to $\cF$, note that
(\ref{b_per}) can be seen as an infinite set of equalities between
polynomials in the three independent variables $E_2$, $E_4$,
$E_6$. Setting these to zero, the integral on the right hand side can be evaluated
in a large $a$ expansion to give
\ba
\oint_{B_+} \sqrt{ \wp - 2 \pi i \partial_\tau \cF - \epsilon_1 \cW'' -\epsilon_1^2\, \frac{\wp}{4} } \,dz |_{E_i=0} &=&  \oint_{B_+} \sqrt{ (1 - \frac{\epsilon_1^2}{4} )\wp + (2 \pi i \, a)^2 }\,dz |_{E_i=0} \nn\\
&=& 2 \pi i\, a \left( \tau + \frac{1}{2} \frac{1}{(2 \pi i\, a)^2} (1 - \frac{\epsilon_1^2}{4} ) 2 \pi i \right)  \nn \\
&=& 2\pi i \, a\tau + \frac{1}{2a} (1 - \frac{\epsilon_1^2}{4} ) \,.
\ea
We have used that $\cW''|_{E_i=0}$ is a formal power series with coefficients in $\wp^2 \IC[\wp] \oplus \wp'\IC[\wp]$, that $\oint_B \wp \, dz |_{E_i=0} = 2 \pi i$, $\oint_B \wp^n \, dz |_{E_i=0} = 0$ for $n>1$ \cite{Grosset}, and that $\partial_\tau \cF |_{E_i=0} = - 2 \pi i \, a^2$. If we hence choose the integration constant in passing from $\partial_\tau \cF$ to $\cF$ such that
\be
\frac{\partial \cF}{\partial a} |_{E_i=0} = -4\pi i \, a \tau - \frac{1}{a} (1 - \frac{\epsilon_1^2}{4} )  \,,
\ee
we set the $\tau$ independent function $G(a)$ in (\ref{b_per}) to zero, and obtain the relation between the $\epsilon_1$-deformed prepotential and Seiberg-Witten differential in the final form  
\be  \label{fixing_int_const}
\frac{\partial \cF}{\partial a} = - 2 \oint_{B_+} \sqrt{ \wp - 2 \pi i \partial_\tau \cF - \epsilon_1 \cW'' -\epsilon_1^2\, \frac{\wp}{4} } \,dz \,.
\ee
The equality (\ref{fixing_int_const}) permits us to fill a gap in our results in
\cite{KashaniPoor:2012wb}. There, we demonstrated that $\partial_\tau
\cF_n$ are quasi-modular, and we verified experimentally at low $n$
that this property is inherited from $\cF_n$. As derivatives with respect to the variable $a$ do
not interfere with quasi-modularity and the right hand side of
(\ref{fixing_int_const}) is manifestly quasi-modular, the relation (\ref{fixing_int_const}) proves the
quasi-modularity of $\cF_n$ to all orders.

Note that if the functions $\cF$ and $\cW$ were analytic, rather than
formal power series, the right hand side of equation (\ref{b_per}) could be
interpreted as the integral of a meromorphic form over a modified (or
quantum-corrected) Seiberg-Witten geometry, which would depend on the solutions to the equation $\wp=U +
\epsilon_1 \cW'' + \epsilon_1^2 \frac{\wp}{4}$. 

A final remark on the null vector decoupling equation (\ref{diff_equ_psi}) is that we can think of the differential equation as the quantization of a deformed Seiberg-Witten curve with the operator $\partial_z$ and the variable $z$ as canonically conjugate variables. The null vector decoupling equation on the torus can thus be interpreted as the quantum curve annihilating the partition function.

Several other approaches to deformed Seiberg-Witten theory have appeared in the literature. In the topological string setting, the notions of deformed periods and curves elevated to differential operators annihilating the partition function were introduced and studied in \cite{Aganagic:2003qj,Aganagic:2011mi}. For results inspired by the relation between the $\epsilon_2 \rightarrow 0$ limit and integrable models, see \cite{Poghossian:2010pn,Fucito:2011pn}. A matrix model approach is developed in \cite{Marshakov:2010fx,Mironov:2009uv,Bourgine:2012gy,Bourgine:2012bv}.

\subsubsection{Comparing to the proposal in \cite{Alday:2009aq} for the Seiberg-Witten geometry}
Above, we have demonstrated that the differential $\cW'\,dz$, defined
in the semi-classical $\epsilon_2 \rightarrow 0$ limit via
equation (\ref{fact_ansatz}), can be identified with the Seiberg-Witten
differential. Defining the Seiberg-Witten curve by the requirement
that this one-form be single-valued, we obtained the hyperelliptic
equation for the curve to be
\be \label{sw_deg}
t^2 = (\cW')^2= \left( \lim_{\epsilon_2 \rightarrow 0} \frac{d}{dz} \log \langle V_{h_m}(0) V_{h_{(2,1)}}(z) \rangle_\tau \bigg\rvert_{h,h_\pm} \right)^2 \,. 
\ee
The choice of the second internal momentum $h_{\pm}$ simply determines the overall sign of $\cW'$.
The following Seiberg-Witten curve was proposed in \cite{Alday:2009aq}:\footnote{We have adjusted the limit to our parametrization of the variables.}
\be \label{sw_vev_t}
t^2 = \lim_{\epsilon_2 \rightarrow 0} \epsilon_1 \epsilon_2 \frac{\langle T(z) V_{h_m}(0) \rangle_\tau \Big\rvert_h}{\langle V_{h_m}(0) \rangle_\tau \Big\rvert_h }  \,,
\ee
with Seiberg-Witten differential $tdz$.
To compare the two proposals, we can evaluate the right hand side of 
equation (\ref{sw_vev_t}) by invoking the Ward identity \cite{Eguchi:1986sb}:
\begin{eqnarray} \label{ward_id}
\lefteqn{\langle T(z) \prod_{i=1}^n\Vt_i(z_i) \rangle - \langle T \rangle \langle \prod_{i=1}^n \Vt_i(z_i) \rangle =}\\
&=& \sum_{i=1}^n \Big( h_i ( \wp(z-z_i)+ 2 \eta_1)  +(\zeta(z-z_i)+2 \eta_1 z_i) \partial_{z_i}
\Big) \langle \prod_{i=1}^n \Vt_i(z_i) \rangle 
 + 2 \pi i \partial_\tau \langle \prod_{i=1}^n\Vt_i (z_i)\rangle \,.  \nonumber  
\end{eqnarray}
We obtain 
\be
\frac{\langle T(z) \Vt_{h_m}(0) \rangle}{\langle \Vt_{h_m}(0) \rangle} =  h_m ( \wp(z)+ 2 \eta_1)  + 2 \pi i \partial_\tau \log \langle V(0) \rangle + \langle T \rangle \,.  \label{tvev}
\ee
Projecting this equation onto the $h$ channel and substituting our definition of 
the one-point conformal block $\cF$ from equation (\ref{one_point_F})  yields
\be
\lim_{\epsilon_2 \rightarrow 0} \epsilon_1 \epsilon_2  \frac{\langle T(z) V_{h_m}(0) \rangle \Big\rvert_h}{\langle V_{h_m}(0) \rangle \Big\rvert_h}=  \left( \frac{\epsilon_1^2}{4}-m^2 \right) \wp(z)  + 2 \pi i \partial_\tau \cF \,,
\ee
where we have used $\langle T \rangle = 2 \pi i \partial_\tau \log Z$. This is to be compared to the expression (\ref{diff_equ_fact}) for $(\cW')^2$ we obtain by imposing null vector decoupling,
\be
\epsilon_1^2 \left\langle (L_{-2} \Vt_{(2,1)}) (w) V_{h_m}(0)  \right\rangle = \frac{1}{\epsilon_2^2} \left\langle (L_{-1}^2 \Vt_{(2,1)})(w)  \Vt_{h_m}(0)  \right\rangle \,.
\ee
Projecting onto the $h, h_\pm$ channel, dividing both sides of this equation by the two-point block, and taking the $\epsilon_2 \rightarrow 0$ limit yields \cite{KashaniPoor:2012wb}
\be
(\cW')^2 + \epsilon_1 \cW''  = \left( \frac{\epsilon_1^2}{4}-m^2 \right) \wp(z)  + 2 \pi i \partial_\tau \cF \,.
\ee
The proposals in equations (\ref{sw_deg}) and (\ref{sw_vev_t}) hence
yield the same classical Seiberg-Witten curve (defined at
$\epsilon_1=0$). The additional term $\epsilon_1 \cW''$ arising from
(\ref{sw_deg}) enters into the definition of the deformed Seiberg-Witten differential (\ref{def_def_diff}) and is necessary for reproducing the amplitudes
$\cF_n$ expected from gauge theory beyond the lowest order in $\epsilon_1$.

\subsubsection{Bohr-Sommerfeld interpretation}
Our analysis can be cast in the light of a Bohr-Sommerfeld evaluation of the Schr\"odinger-like equation (\ref{diff_equ_psi}) in the $\epsilon_2 \rightarrow 0$ limit,
\be 
\left(-\epsilon_1^2\, \partial_z^2+ V(z)  \right) \Psi(z|\tau) = u \, \Psi(z|\tau) 
\ee 
with 
\be  \label{bsu}
V(z) = (m^2 - \frac{\epsilon_1^2}{4} ) \wp(z) \,, \quad u = 2\pi i \partial_\tau \cF  \,.
\ee
It was pointed out in \cite{Marshakov:2010fx} that the 2d/4d
correspondence permits determining Schr\"odinger equations associated
to a Seiberg-Witten theory via null vector decoupling equations. The
analysis closest in spirit to ours appears in \cite{Mironov:2009uv},
where the sine-Gordon quantum model is used to compute the
$\epsilon_1$-deformed prepotential of pure $SU(2)$ gauge theory. The
authors compute the $A$ and $B$ periods $\Pi_A$ and $\Pi_B$ of the
exact Bohr-Sommerfeld integral ($\lambda$ in the notation above) as
functions of $u$, invert $a = \Pi_A(u)$ to obtain $u(a)$, and then
impose $\frac{\partial F(a | \epsilon_1)}{\partial a} = \Pi_B(u(a))$
to determine $F( a | \epsilon_1)$. They establish to low orders in
$\epsilon_1$ that the $F(a|\epsilon_1)$ thus computed coincides with
the $\epsilon_1$-deformed prepotential, thereby providing
evidence for a claim in \cite{Nekrasov:2009rc}. Given the 2d/4d
correspondence, our analysis above in fact proves to all orders that
this procedure must yield the deformed prepotential: the 2d/4d
correspondence identifies the $\cF$ appearing in (\ref{bsu}) with this
prepotential, while we proved in subsection \ref{mat_bey_leading} that the
$B$ period of $\lambda$ coincides with the $a$-derivative of $\cF$.

\subsection{Transformation properties and S-duality}
The quasi-modular $\tau$-dependence of the coefficients of $\cF$ in a
formal $\epsilon_1$-expansion is clearly a consequence of $S$-duality
in gauge theory, through the 2d/4d correspondence.
Encountering {\it quasi-}modularity in this context may be surprising at first blush, because such forms in fact transform in a rather messy way under the $S$-transformation
of the modular group $SL(2,\IZ)$.

Electromagnetic duality states that the same $\cN=2$ gauge theory
expressed in terms of electric or magnetic variables has infrared couplings
related by $\tau_D^{IR} = -1/\tau^{IR}$. The derivatives of the
corresponding prepotentials, $h=F'(a)$ and $h_D=F'(a_D)$, must hence be
inverse functions of each other, up to a sign (\cite{Seiberg:1994rs}):
$h_D(h(a)) = - a$. Two functions whose derivatives are inverse
functions of each other are themselves related by Legendre transform,
hence $F_D(a_D) = F(a) - a_D a$.

S-duality identifies superconformal theories specified by different ultraviolet
data. In the case of $\cN=2^*$, we will demonstrate, using our conformal
field theory approach, that the
prepotentials of the theories at ultraviolet couplings $\tau$ and $-1/\tau$
behave as $F$ to $F_D$, in the sense that
\be
F(a_D;-1/\tau ) = F(a;\tau) - a_D a  \,.
\ee
We will furthermore prove that this relation persists to all orders in
$\epsilon_1$. To this end, we introduce the notation $\cF(\tau;a)$ and
$\cW(z,\tau;a)$ to indicate the unique solution of the differential
equation (\ref{diff_equ_fact}) satisfying the boundary condition
\be  \label{bc_not}
\oint_{A_+} \cW'(z,\tau;a) = 2 \pi i a
\ee
and
\be  \label{princ_int}
\oint_{A_+} \sqrt{ (1- \frac{\epsilon_1^2}{4}) \wp(z,\tau) - 2\pi i \partial_\tau \cF(\tau;a) - \epsilon_1 \partial_z^2 \cW(z,\tau;a)}  dz = 2 \pi i a \,.
\ee
Note that as long at the $\partial_\tau$ derivative on $\cF$ is
defined as acting only on the first argument, we are also entitled to
endow $a$ with $\tau$-dependence. Let us now consider
\ba
2 \pi i \,a_D(-1/\tau) &=& \oint_{B_+} \sqrt{ (1- \frac{\epsilon_1^2}{4}) \wp(z,-\frac{1}{\tau}) - 2\pi i \partial_1 \cF(-\frac{1}{\tau};a) - \epsilon_1 \partial_z^2 \cW(z,-\frac{1}{\tau};a)}  dz \nn \\
&=& \int_0^{-\frac{1}{\tau}}  \sqrt{ (1- \frac{\epsilon_1^2}{4}) \wp(z,-\frac{1}{\tau}) - 2\pi i \partial_1 \cF(-\frac{1}{\tau};a) - \epsilon_1 \partial_z^2 \cW(z,-\frac{1}{\tau};a)} dz \nn\\
&=& \frac{1}{\tau} \int_0^{-1}  \sqrt{( 1- \frac{\epsilon_1^2}{4}) \wp(\frac{z}{\tau},-\frac{1}{\tau}) - 2\pi i \partial_1 \cF(-\frac{1}{\tau};a) - \epsilon_1 \partial_1^2 \cW(\frac{z}{\tau},-\frac{1}{\tau};a)} dz \nn \\
&=& -\int_0^{1}  \sqrt{( 1- \frac{\epsilon_1^2}{4}) \wp(z,\tau) - 2\pi i \partial_\tau \cF(-\frac{1}{\tau};a) - \epsilon_1 \partial_z^2 \cW(\frac{z}{\tau},-\frac{1}{\tau};a)} dz \,. \nn\\ \label{trans1}
\ea
In the last line, we have used the fact that both $\wp$ and
$\cW''(z/\tau, -1/\tau)$ are invariant under $z \rightarrow z+1$, the
latter via (\ref{form_W}).  By considering the asymptotic expansion,
and the lowest order in $\epsilon_1$ explicitly, we conclude that
\be
2\pi i \partial_\tau \cF(-\frac{1}{\tau};a) + \epsilon_1 \partial_z^2 \cW(\frac{z}{\tau},-\frac{1}{\tau};a) = 2\pi i \partial_1 \cF(\tau;-a_D(-1/\tau)) + \epsilon_1 \partial_z^2 \cW(z,\tau;-a_D(-1/\tau))  \,.
\ee
Calculating the $A_+$ period of both
sides and invoking (\ref{form_W}) finally yields 
\be
\partial_\tau \cF(-\frac{1}{\tau};a) = \partial_1 \cF(\tau;a_D(-1/\tau)) \,,
\ee
or equivalently
\be \label{trans_dtau_F}
\partial_\tau \cF(\tau;a) = \partial_\tau \cF(-1/\tau;a_D(\tau)) \,,
\ee
with the $\tau$-derivative on the right hand side only acting on the first argument of $\cF$. We have here used the fact that $\cF$ is an even function of $a$. Since $\partial_z \cW$ is odd under $(z,a) \rightarrow -(z,a)$ \cite{KashaniPoor:2012wb}, this allows us to conclude that
\be \label{trans_W}
\partial_z^2 \cW(z,\tau;a) = \partial_z^2 \cW(\frac{z}{\tau},-\frac{1}{\tau};a_D(\tau)) \,.
\ee
To integrate (\ref{trans_dtau_F}), we need to pass from partial to total $\tau$-derivatives. Assuming that $a$ is $\tau$ independent, this is
\be
\frac{d}{d \tau} \cF(\tau;  a) = \frac{d}{d \tau} \cF(-1/\tau;  a_D)  - \partial_2 \cF(-1/\tau;  a_D) \frac{ d a_D}{d\tau} \,.
\ee
Starting from (\ref{fixing_int_const}), a calculation very similar to (\ref{trans1}) invoking  (\ref{trans_dtau_F}) and (\ref{trans_W}) yields
\ba  
\partial_2 \cF(-1/\tau;a_D)
 &=& -2 \oint_{B_+} \sqrt{ ( 1- \frac{\epsilon_1^2}{4})\wp (z,-\frac{1}{\tau}) - 2 \pi i \partial_1 \cF(-\frac{1}{\tau};a_D) - \epsilon_1 \partial_1^2\cW(z,-\frac{1}{\tau};a_D)  } \,dz    \nn  \\
&=& -2 \int_0^{-\frac{1}{\tau}} \sqrt{ ( 1- \frac{\epsilon_1^2}{4})\wp (z,-\frac{1}{\tau}) - 2 \pi i \partial_1 \cF(-\frac{1}{\tau};a_D) - \epsilon_1 \partial_1^2 \cW(z,-\frac{1}{\tau};a_D)  } \,dz    \nn  \\
&=& -\frac{2}{\tau} \int_0^{-1} \sqrt{ ( 1- \frac{\epsilon_1^2}{4})\wp (\frac{z}{\tau},-\frac{1}{\tau}) - 2 \pi i \partial_1 \cF(-\frac{1}{\tau};a_D) - \epsilon_1 \partial_1^2 \cW(\frac{z}{\tau},-\frac{1}{\tau};a_D)  } \,dz    \nn  \\
&=& 2 \int_0^{1} \sqrt{ ( 1- \frac{\epsilon_1^2}{4})\wp (z,\tau) - 2 \pi i \partial_\tau \cF(-\frac{1}{\tau};a_D) - \epsilon_1 \partial_z^2 \cW(\frac{z}{\tau},-\frac{1}{\tau};a_D)  } \,dz    \nn  \\
&=& 2 \int_0^{1} \sqrt{ ( 1- \frac{\epsilon_1^2}{4})\wp (z,\tau) - 2 \pi i \partial_\tau \cF(\tau;a) - \epsilon_1 \partial_z^2 \cW(z,\tau;a)  } \,dz    \nn  \\
&=& 4 \pi i \,a \,.  \label{divFD}
\ea
Hence,
\be  \label{Legendre}
\cF(-1/\tau;a_D) = \cF(\tau,a) + 4 \pi i \,a a_D + C \,.
\ee
Taking the total $a$-derivative on both sides demonstrates that the constant
$C$ is independent of $a$ as well as of $\tau$. Explicit
computation shows that the contributions to the constant stem from
orders 0 and 1 in $\epsilon_1^2$, and that $C= -\frac{1}{2} \pi
i ( 1- \frac{\epsilon_1^2}{4})$.

\section{Exact conformal field theory methods}
\label{smove}
In the previous section, we computed the monodromy of the two-point
conformal block (\ref{2pt_deg}) around the $B$-period of the torus and
the transformation properties of the one-point conformal block
(\ref{one_point}) under the S-transformation $\tau \rightarrow -
1/\tau$ by analyzing the null vector decoupling equation in the
semi-classical limit. Both computations can be
performed exactly in conformal field theory. By re-deriving our results from the semi-classics of these exact relations, we demonstrate that it is consistent to take the semi-classical approximation already at the level of the null vector decoupling equations. Extracting the gauge theory/topological string theory amplitudes from these exact results is a first step towards moving beyond perturbation theory on this side of the 2d/4d correspondence.

\subsection{The $B$-monodromy from braiding}
We can compute the $A$-monodromy of the two-point toroidal block in the position of the degenerate operator from the operator product expansions 
\ba
\phi_{h_{(2,1)}}(z) | a \rangle &=& \phi_{h_{(2,1)}}(z)  \phi_a(0) |0 \rangle \nn \\
&=& \left( C_{h_{(2,1)},h_a}^{h_+} z^{h_+ - h_{(2,1)} - h_a} \left( \phi_{h_+}(0) + \ldots \right) +  C_{h_{(2,1)},h_a}^{h_-} z^{h_- - h_{(2,1)} - h_a} \left( \phi_{h_-}(0) + \ldots \right) \right) | 0 \rangle \,, \nn
\ea
or
\ba
\langle a|  \phi_{h_{(2,1)}}(z) &=& \lim_{w \rightarrow \infty} w^{2h_a} \langle 0 | \phi_{h_a}(w) \phi_{h_{(2,1)}}(z) \nn\\
&=& \lim_{\tilde{w} \rightarrow 0} \tz^{2h_{(2,1)}} \langle 0 | \tphi_{h_a}(\tw) \tphi_{h_{(2,1)}}(\tz) \nn\\
&=& \langle 0 | \Big(C_{h_a,h_{(2,1)}}^{h_+} (-\tz)^{h_+ + h_{(2,1)} - h_a} \left( \phi_{h_+}(0) + \ldots \right) \nn \\
&& \quad+  C_{h_a,h_{(2,1)}}^{h_-} (-\tz)^{h_- + h_{(2,1)} - h_a} \left( \phi_{h_-}(0) + \ldots \right) \Big) \,, \nn
\ea
by circling the origin or infinity respectively, obtaining the same two monodromies in the semi-classical limit. This avenue of computation is available as the operator product expansion remains valid along the entire path associated to the monodromy. By contrast, the $B$-monodromy requires exchanging the order of the two operator insertions along the monodromy path. Its computation hence requires invoking braiding matrices. These relate the conformal blocks
\ba
\begin{aligned}
\xymatrix@!0@M0pc@W0pc@H0pc{  & &  & & & \\
       &   _{i} \ar@{-}'[rrrr]
& \ar@{-}[u]^>>{j} &  _{p}   &\ar@{-}[u]^>>{k} & _{l}
\\
& & & & &
}
\end{aligned} 
&=&
B^\eta_{pq} 
\begin{bmatrix}
j & k \\
i & l
\end{bmatrix}
\hspace{-0.8cm}
\begin{aligned}
\xymatrix@!0@M0pc@W0pc@H0pc{  & &  & & & \\
       &  _{i}  \ar@{-}'[rrrr]_{q}
& \ar@{-}[u]^>>{k} &     &\ar@{-}[u]^>>{j} &  _{l}
\\
& & & & &
}
\end{aligned}
\ea
The index $\eta =\pm$ indicates the sense of the braiding. It will not play a role in the following. As usual, we glue the two ends
of the diagram to obtain a torus conformal block by inserting a
translation operator $q^H$ and summing over initial and final
states. By choosing these states in an eigenbasis of $H$, we see that
the braiding matrix is not affected by the insertion of this operator.

We can relate the two-point function evaluated at arguments $z$ and $z+\tau$ via the following sequence of manipulations:
\ba
Z(a,a_+;z-w) &=& \langle a | q^H \Phi_{h_m}(w) | a_+ \rangle \langle a_+ | \Phi_{h_{(2,1)}}(z) | a \rangle \\
&=& \langle a_+ | \Phi_{h_{(2,1)}}(z) | a \rangle  \langle a | q^H \Phi_{h_m}(w) | a_+ \rangle \\
&=& \langle a_+ | \Phi_{h_{(2,1)}} (z) q^H | a \rangle  \langle a | \Phi_{h_m}(w) | a_+ \rangle \\
&=& \langle a_+ | q^H \Phi_{h_{(2,1)}} (z+\tau)  | a \rangle  \langle a | \Phi_{h_m}(w) | a_+ \rangle \\
&=& \sum_{a'=a,a_{++}} B^{\eta}_{a a'}
\begin{bmatrix}
  -b/2 & \alpha_m \\
   \alpha_+ & \alpha_+ \\
\end{bmatrix}
\langle a_+ | q^H \Phi_{h_m}(w)  | a' \rangle  \langle a' |  \Phi_{h_{(2,1)}} (z+\tau) | a_+ \rangle  \nn 
\ea
with $a_+=a+\frac{\epsilon_2}{2}$, $a_{++}=a+\epsilon_2$, $\alpha_+ = \frac{Q}{2} + \frac{a_+}{\sqrt{\epsilon_1 \epsilon_2}}$, $\alpha_m = \frac{Q}{2} + \frac{m}{\sqrt{\epsilon_1 \epsilon_2}}$ . The notation here is that repeated states imply a sum over the descendants of the indicated primaries. Diagrammatically,
\ba
\begin{aligned}
\xymatrix@!0@M0pc@W0pc@H0pc{  & &  & & & \\
       &   & \ar@{-}`[rrr]_<<{w}_(.8){z}
`[rrrd]  `[ld]^{a}  `[l]  []   \ar@{-}[u]^{\alpha_m}
& ^{a_+} &   \ar@{-}[u]^{-\frac{b}{2}}  &
\\
& & & & &
}
\end{aligned}
&=& \hspace{-0.8cm}
\begin{aligned}
\xymatrix@!0@M0pc@W0pc@H0pc{  & &  & & & \\
       &   & \ar@{-}`[rrr]_<<{z+\tau}_(.8){w}
`[rrrd] `[ld]^{a_+}  `[l]  []   \ar@{-}[u]^{-\frac{b}{2}}
& ^a  &   \ar@{-}[u]^{\alpha_m}  &
\\
& & & & &
}
\end{aligned} \\
&=& B^\eta_{aa} \hspace{-0.8cm}
\begin{aligned}
\xymatrix@!0@M0pc@W0pc@H0pc{  & &  & & & \\
       &   & \ar@{-}`[rrr]_<<{w}_(.8){z+\tau}
`[rrrd] `[ld]^{a_+}  `[l]  []   \ar@{-}[u]^{\alpha_m}
& ^a  &   \ar@{-}[u]^{-\frac{b}{2}}  &
\\
& & & & &
}
\end{aligned}
+
B^\eta_{aa_{++}} \hspace{-0.8cm}
\begin{aligned}
\xymatrix@!0@M0pc@W0pc@H0pc{  & &  & & & \\
       &   & \ar@{-}`[rrr]_<<{w}_(.8){z+\tau}
`[rrrd] `[ld]^{a_+}  `[l]  []   \ar@{-}[u]^{\alpha_m}
& ^{a_{++}}  &   \ar@{-}[u]^{-\frac{b}{2}}  &
\\
& & & & &
}
\end{aligned} \nn
\ea

 To justify the manipulations, we assume that an orthonormal basis for each level has been introduced. We can obtain braiding from fusion matrices via \cite{Moore:1988qv}
\be
F_{a a'} 
\begin{bmatrix}
  \a_2 & \a_3 \\
   \a_1 & \a_4 \\
\end{bmatrix}
= 
e^{- i \phi(\eta) } B^\eta_{a a'} 
\begin{bmatrix}
  \a_2 & \a_4 \\
   \a_1 & \a_3 \\
\end{bmatrix} \,,
\ee
with $\phi(\eta) = \eta  \pi ( \Delta_{\a_1} + \Delta_{\a_3} - \Delta_a - \Delta_{a'} )$. With the fusion matrices as derived in \cite{Alday:2009fs}, we arrive at
\ba
B^\eta_{a_{1-} a_{1-}} 
\begin{bmatrix}
 -b/2 & \a_2 \\
  \a_1 & \a_1 \\
\end{bmatrix}
&=& e^{i\phi(\eta)} \frac{ \Gamma[(2\a_1 - b)b] \Gamma[(Q- 2 \a_1) b]}{ \Gamma[( \a_2 - \frac{b}{2})b] \Gamma[1-\a_2 b + \frac{b^2}{2}]}  \,, \\
B^\eta_{a_{1-} a_{1+}} 
\begin{bmatrix}
 -b/2 & \a_2 \\
  \a_1 & \a_1 \\
\end{bmatrix}
&=& e^{i\phi(\eta)} \frac{ \Gamma[(2\a_1 - b)b] \Gamma[-(Q- 2 \a_1) b]}{ \Gamma[(2\a_1 - \a_2 - \frac{b}{2})b] \Gamma[(2 \a_1+\a_2 - \frac{b}{2}-Q)b ]} \,,
\ea
with $\alpha_i = \frac{Q}{2} + \frac{a_i}{\sqrt{\epsilon_1 \epsilon_2}}$. Setting $\a_1 = \alpha_+$, $\a_2 = \alpha_m$, this yields
\ba
B^\eta_{a a} &=& e^{i\phi_1(\eta)} \frac{ \Gamma[ -\frac{2a+\epsilon_2}{\epsilon_1}   ]\Gamma[ 1+\frac{2a+\epsilon_2}{\epsilon_1}  ] }{ \Gamma[\frac{1}{2} - \frac{m}{\epsilon_1} ] \Gamma [ \frac{1}{2} + \frac{m}{\epsilon_1} ] }  \,, \\
B^\eta_{a a_{++}} &=& e^{i\phi_2(\eta)} \frac{ \Gamma[  \frac{2a+\epsilon_2}{\epsilon_1}    ]\Gamma[  1+\frac{2a+\epsilon_2}{\epsilon_1}   ] }{ \Gamma[ \frac{1}{2} + \frac{2a-m+\epsilon_2}{\epsilon_1} ] \Gamma [ \frac{1}{2} + \frac{2a+m+\epsilon_2}{\epsilon_1}   ] }  \,,
\ea
where $\phi_1= - \eta \pi \frac{4 a + \epsilon_2}{2 \epsilon_1}$ and $\phi_2 = \eta \pi  \frac{\epsilon_2}{2 \epsilon_1}$.
Using
\be
\Gamma(\frac{1}{2}+x) \Gamma(\frac{1}{2} -x) = \frac{\pi}{\cos \pi x}  \,,
\ee
\be
\Gamma(1+ix)\Gamma(1-ix) = \frac{ \pi x}{\sinh \pi x} \,, \quad x \in \IR \,,
\ee
and
\be
\lim_{|z| \rightarrow \infty }\frac{\Gamma(z+a)}{\Gamma(z)} e^{-a \log z} =1  \,,
\ee
and setting $a = i \alpha$, $m = i \mu$, $\alpha,\mu \in \IR$, we obtain
\be
B^\eta_{a a} \xrightarrow[\epsilon_2 \rightarrow 0]{}  \quad  -i e^{- \eta i \pi \frac{2 \alpha}{\epsilon_1}} \,\frac{\cosh \pi \frac{\mu}{\epsilon_1}}{\sinh \pi \frac{2 \alpha}{\epsilon_1}} \quad\xrightarrow[\alpha, \mu \rightarrow \infty]{} \quad -i e^{- \eta i \pi \frac{2 \alpha}{\epsilon_1}} \, e^{-\frac{2\pi \alpha}{\epsilon_1} ( 1- \frac{ \mu}{2 \alpha})} \quad\xrightarrow[\alpha > \mu]{} 0
\ee
and
\be
B^\eta_{a a_{++}}  \xrightarrow[\epsilon_2 \rightarrow 0]{}  \quad \frac{ \Gamma[  \frac{2a}{\epsilon_1}    ]\Gamma[  \frac{2a}{\epsilon_1}   ] }{ \Gamma[ \frac{1}{2} + \frac{2a-m}{\epsilon_1} ] \Gamma [ \frac{1}{2} + \frac{2a+m}{\epsilon_1}   ] }  e^{(\frac{1}{2}+\frac{m}{\epsilon_1}) \log \frac{2a}{\epsilon_1}}e^{(\frac{1}{2}-\frac{m}{\epsilon_1}) \log \frac{2a}{\epsilon_1}} \xrightarrow[|a| \rightarrow \infty]{} 1 \,.
\ee
In the limit we are considering, we thus obtain the following relation between conformal blocks:
\be
Z(a,a_+;z-\tau) \sim Z(a_{+},a_{++};z) \,.
\ee
Defining $Z(a;z) := Z(a,a_+;z)$, dividing the above equation by $Z(a;z)$ on both sides, and making the semiclassical ansatz $Z = \exp \frac{1}{\epsilon_1 \epsilon_2} \cF(a) + \frac{1}{\epsilon_1} \cW(z;a)$, we arrive at
\ba
\lim_{\epsilon_2 \rightarrow 0} \log \frac{Z(a;z-\tau) }{Z(a;z)} &=& \frac{1}{\epsilon_1} \left(\cW(z-\tau) - \cW(z) \right) \\
&=& \lim_{\epsilon_2 \rightarrow 0} \log \frac{Z(a_+;z) }{Z(a;z)}  \\
&=& \lim_{\epsilon_2 \rightarrow 0} \frac{1}{\epsilon_1 \epsilon_2} \left( \cF(a+\frac{\epsilon_2}{2}) -  \cF(a) \right) \\
&=&  \frac{1}{2\epsilon_1}  \partial_a \cF(a) \,,
\ea
thus reproducing the relation (\ref{fixing_int_const}). 

A related line of reasoning, invoking Verlinde operators, appears in \cite{Alday:2009fs}.

\subsection{The semi-classical S-move kernel} \label{semi_S}
The S-move kernel relates the one-point conformal blocks with
Teichm\"uller parameter $\tau$ and $-1/\tau$. It is naturally defined in conventions in which the three-point function
contribution in (\ref{one_point}) is absorbed in the conformal blocks. Denoting the rescaled one-point blocks as $F_{h_m}^{h(p)}$, the one-point function is given by \cite{Teschner:2003at,Vartanov:2013ima} 
\be
\langle \Vt_{{h}_m} \rangle_\tau =  \int_0^\infty dp \,\mu(p) F_{h_m}^{h(p)}(\tau) \bar{F}_{h_m}^{h(p)}(\tau)  \,,
\ee
where the weight $h(p)$ is parametrized as $h(p) = (\frac{Q}{2} + ip) (\frac{Q}{2} - ip)$ and $\mu(p)$ is the measure factor
\be
\mu (p)  = 4  \sinh 2 \pi  p b \sinh 2 \pi b^{-1} p \, .
\label{measurefactor}
\ee
The integral kernel implementing the S-transformation for the block $F_{h_m}^{h(p)}(\tau)$ via
\be  \label{S_move}
F_{h_m}^{h(p_2)}(\tau) = \int_{0}^\infty d p_1 \, \mu(p_1) S_{p_2 p_1} (p_m) F_{h_m}^{h(p_1)} (-\frac{1}{\tau}) 
\ee
is given by \cite{Teschner:2003at}
\begin{eqnarray}
S_{p_2 p_1} (p_m) &=& \frac{2^{\frac{3}{2}}}{s_b(p_m)} \int_{\mathbb{R}}
dr \prod_{\epsilon=\pm} \frac{s_b(p_1+ \frac{1}{2} (p_m+ i \frac{Q}{2})+ \epsilon r)}{s_b(p_1- \frac{1}{2} (p_m+ i \frac{Q}{2})+ \epsilon r)}
e^{4 \pi i p_2 r}  \,,
\label{Skernel}
\end{eqnarray}
where the function $s_b$ is defined by
\be
\log s_b (x) = \frac{1}{i} \int_0^\infty \frac{dt}{t} ( \frac{\sin 2 x t}{2 \sinh bt \sinh b^{-1} t}- \frac{x}{t})
\, .
\ee
The kernel is invariant under $p_1 \rightarrow -p_1$ as well as $p_2 \rightarrow -p_2$, the latter due to the functional relation
\be
s_b(x) s_b(-x) = 1 \,.
\ee
We would like to compare the exact result (\ref{S_move}) to the transformation properties of the conformal block $\cF$ that we derived in
equation (\ref{Legendre}). The identification of parameters\footnote{As the kernel is independent of $\tau$, the relation (\ref{S_move}) remains valid with $\tau$ replaced by $-1/\tau$. There is hence no natural distinction between $a$ and $a_D$ in this context.}
\begin{eqnarray}
p_1 &=& - i \frac{a_1}{ \sqrt{\epsilon_1 \epsilon_2}}  \,, \\
\nonumber \\
p_2 &=& -i \frac{a_2}{ \sqrt{\epsilon_1 \epsilon_2}}   \,, \\
\nonumber \\
p_m &=&  -i \frac{m}{ \sqrt{\epsilon_1 \epsilon_2}}  \,,
\label{dico}
\end{eqnarray}
shows that the $\epsilon_2 \rightarrow 0$ limit corresponds to the limit in which all momenta are taken large.
In this limit, the shift relation
\be
s_b (x-ib) = 2 \cosh \pi b x \,s_b(x) 
\ee
gives rise to a first order differential equation for the function $\log s_b$, 
which we can integrate using one special value (e.g. $s_b(0)=1$), giving rise to the approximation
\begin{eqnarray}
\lim_{b \rightarrow 0} \log s_b(x) & \approx & \frac{i}{b} \int_0^{x} dx' \log 2 \cosh \pi b x' \, .
\end{eqnarray}
We obtain an error estimate for this approximation in appendix \ref{app_sb}. The S-move kernel in the semi-classical $b \rightarrow 0$ limit is thus approximated by
\be
S_{p_2 p_1} (p_m)  \approx  \frac{2^{\frac{3}{2}}}{ s_b(p_m)}  \int_{\mathbb{R}}
dr 
\exp \left[ 4 \pi i p_2 r+  {\frac{i}{b} \sum_{\delta,\epsilon = \pm} \delta \int_0^{p_1 + \frac{\delta}{2}(p_m+i \frac{Q}{2})+ \epsilon r} \log (2 \cosh \pi b y') dy'
  } \right] \nn \, .\\ \label{sk_approx}
\ee
Introducing the variables $\alpha_1 =- i a_1$, $\alpha_2 = - i a_2$, $\mu =- i m$, we obtain upon the substitution $r \rightarrow \sqrt{\epsilon_1 \epsilon_2} r$
\be
S_{p_a p_b} (p_e)  \approx  \frac{2^{\frac{3}{2}}}{ \sqrt{\epsilon_1 \epsilon_2}s_b(p_e)} \int_{\mathbb{R}}
dr 
\exp \left[ \frac{1}{\epsilon_2} \left(\frac{ 4 \pi i\alpha_2 r}{\epsilon_1} +  i \sum_{\delta,\epsilon = \pm} \delta \int_0^{\alpha_1 + \frac{\delta}{2}( \mu + i \frac{\epsilon_1 + \epsilon_2}{2}) + \epsilon r} \log (2 \cosh \frac{\pi y}{\epsilon_1}) dy  \right) \right] \nn \, .
\ee
We will evaluate this expression in a saddle point approximation in the limit $\epsilon_2 \rightarrow 0$.
The saddle points of the exponent satisfy the equation \cite{Hadasz:2006vs} (see also \cite{Dimofte:2011jd})
\be
1= e^{\frac{ 4 \pi \alpha_2 }{\epsilon_1} } \prod_{\delta,\epsilon = \pm} \left[ \cosh \left( \frac{\pi}{\epsilon_1} (\alpha_1 + \frac{\delta}{2}( \mu + i \frac{\epsilon_1 + \epsilon_2}{2}) + \epsilon r)\right) \right]^{\delta \epsilon} \,.
\ee
By invoking
\be
\cosh \left(\frac{a+b}{2} + i \frac{\pi}{4} \right) \cosh \left(\frac{b-a}{2} + i \frac{\pi}{4} \right) = \frac{1}{2} \left( \cosh a + i \sinh b \right) \,,
\ee
this equation can be put in the form \cite{Hadasz:2006vs}
\be
e^{\frac{- 4 \pi \alpha_2 }{\epsilon_1} }= \frac{\cosh \frac{2 \pi \alpha_1}{\epsilon_1} + i \sinh \frac{\pi (2r + \mu)}{\epsilon_1}}  {\cosh \frac{2 \pi \alpha_1}{\epsilon_1} - i \sinh \frac{\pi (2r - \mu)}{\epsilon_1}}  \,,
\ee
yielding
\be
\cosh \frac{2 \pi r}{\epsilon_1} = \pm i \sinh \frac{ \pi (2\alpha_1 \mp \mu)}{\epsilon_1} + O (e^{\frac{- 4 \pi |\re \,\alpha_{2} |}{\epsilon_1} }) \quad \rm{for} \quad \re \, \alpha_2 \rightarrow \pm \infty \,,
\ee
and thus
\be
\pm r = \alpha_1 \mp \frac{1}{2} \left(\mu - i \frac{\epsilon_1}{2} \right) + i k \epsilon_1 + O (e^{\frac{- 4 \pi |\re \,\alpha_{2}| }{\epsilon_1} })\,, \quad k \in \IZ \,,\quad \rm{for} \quad \re \, \alpha_2 \rightarrow \pm \infty \,,
\ee
where the $\pm$ on the left hand side in the last equation is not correlated with the sign of $\re\, \alpha_2$. Let us consider the four saddle points closest to the integration path of $r$, which runs along the real axis. Of these, two are zeros of the integrand of (\ref{Skernel}), hence do not correspond to maxima of the real part of the exponential in (\ref{sk_approx}). The other two lie on poles of the integrand of (\ref{Skernel}). The integrand evaluated at these yields\footnote{We are shifting the integration contour to run through the saddle points. As these coincide with the poles of the integrand to $O (e^{\frac{- 4 \pi |\re \,\alpha_{\tiny{D}}| }{\epsilon_1} })$, we evaluate the principal value contribution to the integral around these poles. Note that whether the integration path runs above or below the pole is irrelevant for our computation, as the difference between the two is cancelled in relating the integral on the shifted contour to the original integral. 
}
\ba
\mu(p_1) S_{p_2 p_1}(p_m) \approx \left(e^{\frac{2 \pi \alpha_1}{\epsilon_2}} - e^{-\frac{2 \pi \alpha_1}{\epsilon_2}} \right)  \frac {s_b( \frac{\pm 2 \alpha_1 + i \frac{\epsilon_1}{2}}{\sqrt{ \epsilon_1 \epsilon_2}} ) }{s_b( \frac{\pm 2 \alpha_1 - \mu}{\sqrt{ \epsilon_1 \epsilon_2}} )} &&\!\!\!\!\!\!\!\!\!\cosh \frac{4 \pi i \alpha_2(\alpha_1 \mp \frac{1}{2}(\mu - i \frac{\epsilon_1}{2}))}{\epsilon_1 \epsilon_2}\nn \\
&& \quad \rm{for} \quad \re \, \alpha_2 \rightarrow \pm \infty \,.
\ea
Using this approximation of the S-kernel, a saddle point approximation of the integral (\ref{S_move}) over $p_1$ yields
\be  \label{p1_saddle_point}
\partial_{a_1} \cF_r(a_1,-\frac{1}{\tau}) \approx \mp 4 \pi i a_2 \quad \rm{for} \quad \re \, \alpha_2 \rightarrow \pm \infty \,,
\ee
where $\cF_r$ denotes the amplitude associated to the rescaled conformal block $F_{h_m}^{h(p)}$ introduced above.
To leading order, recalling $\cF_0(a_1,\tau) \approx -2 \pi i a_1^2 \tau$, the relation (\ref{p1_saddle_point}) evaluates to
\be
-a_1 \frac{1}{\tau} = \pm a_2  \quad \rm{for} \quad \re \, \alpha_2 \rightarrow \pm \infty \,.
\ee
Given the integration region $i \IR^+$ for $a_1$, the saddle point which contributes hence depends on the sign of $\re \, \tau$. For $\mathrm{sign} (\re \, \tau)= \pm 1$, we obtain
\ba  
\cF_r(\tau, a_2) &\approx& \pm 4 \pi i \alpha_1 \alpha_2 +2 \pi i (\alpha_1 + \alpha_2) (\mu - i\frac{\epsilon_1}{2}) \pm \frac{\pi i}{2} (\mu^2 + \frac{\epsilon_1^2}{4})   + \cF_r(-\frac{1}{\tau},a_1) \nn\\
&=& \mp 4 \pi i a_1 a_2 -2 \pi i (a_1 + a_2) (m+\frac{\epsilon_1}{2}) \mp \frac{\pi i}{2} (m^2 - \frac{\epsilon_1^2}{4})  + \cF_r(-\frac{1}{\tau},a_1) \,,  \nn \\  \label{legendre_from_kernel}
\ea
where we have used
\be
s_b(y)  \approx  e^{\pm \frac{\pi i }{2} ( y^2+ \frac{1}{12b^2} )} \quad \mbox{for} \quad \re(y) \rightarrow \pm \infty \,,
\ee
see appendix \ref{app_sb}. Matching to (\ref{Legendre}) requires the identification $a_2 = a$, $a_1 = - a_D$ at $\mathrm{sign} (\re \, \tau)= 1$. The terms linear in $a_i$ in the exponential on the right hand side of  (\ref{legendre_from_kernel}) cancel the rescaling of the conformal blocks. While the dependence on the sign of $\re \, \tau$ is unusual, note that choosing the fundamental domain of $\tau$ such that this sign is fixed, it is flipped by $\tau \rightarrow -\frac{1}{\tau}$. It can hence serve to distinguish between electric and magnetic variables, which is indeed the role it is playing in equations (\ref{p1_saddle_point}) and (\ref{legendre_from_kernel}). Recall that already in the derivation of (\ref{Legendre}), the argument $-a_D$ in $\cF(-\frac{1}{\tau},-a_D)$ appeared in the Legendre transform. As $\cF$, in contrast to $\cF_r$, is an even function in this argument, this distinction was not relevant there. We have thus reproduced the Legendre transform relating $\cF$ at $\tau$ and $-\frac{1}{\tau}$ from the S-kernel.

\section{Conclusions}
\label{conclusions}
We have seen how $\epsilon_1$-deformed Seiberg-Witten relations of $\cN=2^*$ gauge theory arise
naturally within conformal field theory in the context of the 2d/4d
correspondence. In particular, we obtained an $\epsilon_1$-deformed Seiberg-Witten differential whose $B$-period evaluated on the classical Seiberg-Witten curve gives rise to the derivative of the deformed prepotential. These tools allowed us to prove
quasi-modularity of the coefficients of the prepotential from
first principles. In the process, we provided a proof of the Matone relation for $\cN=2^*$ theory. We also demonstrated how the deformed relations can be extracted from the semi-classics of exact
conformal field theory quantities. 

An important problem for future
study is moving beyond leading order in the deformation parameter
$\epsilon_2$. Aside from recovering all amplitudes $F^{(n,g)}$ from
within conformal field theory, it would be important to understand
what further modification of the Seiberg-Witten data is necessary to
incorporate these additional corrections. To lift the analysis from
gauge theory geometrically engineered within string theory to the
topological string proper, it would be interesting to formulate and
study a $q$-deformed version of the null vector decoupling equations. 
Finally, the exact results in conformal field theory which
complete relations among the $F^{(n,g)}$ non-perturbatively beg to be
interpreted from a gauge theory/topological string theory perspective.

\section*{Acknowledgments}
We would like to thank Francisco Morales and J\"org Teschner for useful conversations. Our work is supported in part by ANR-grant ANR-13-BS05-0001.

\appendix

\section{The function $s_b$}  \label{app_sb}
The function $s_b$ has the integral representation
\be
\log s_b (x) = \frac{1}{i} \int_0^\infty \frac{dt}{t} \left( \frac{\sin 2 x t}{2 \sinh bt \sinh b^{-1} t}- \frac{x}{t}  \right)
\, .
\ee
We can evaluate the $x$-derivative of this integral by the method of residues:
\ba
\frac{d}{dx} \log s_b(x) &=& \frac{1}{i} \int_0^\infty dt \left( \frac{\cos 2 x t}{ \sinh bt \sinh b^{-1} t}- \frac{1}{t^2} \right) \\
&=& \frac{1}{i} \mathrm{P} \!\!\!\int_{-\infty}^\infty dt \left( \frac{e^{2i x t}}{ 2\sinh bt \sinh b^{-1} t}- \frac{1}{2t^2} \right) \\
&=& \frac{1}{i} ( \int\displaylimits_{\includegraphics[width=0.9cm]{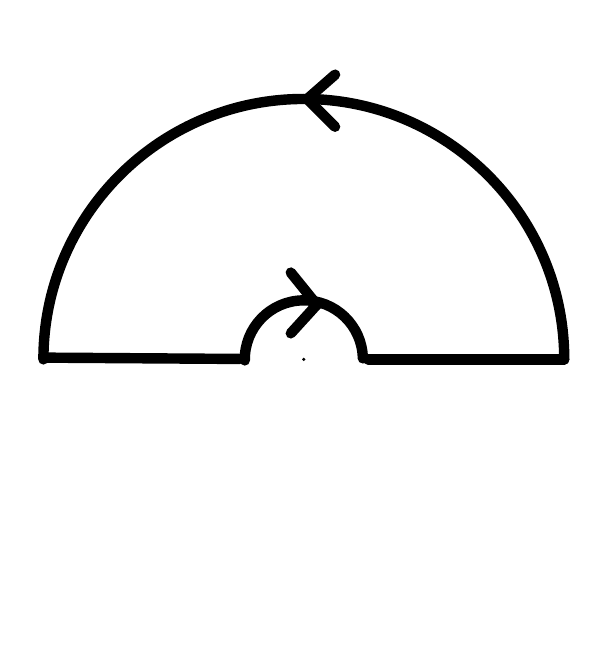}}{}  -\int\displaylimits_{\includegraphics[width=0.9cm]{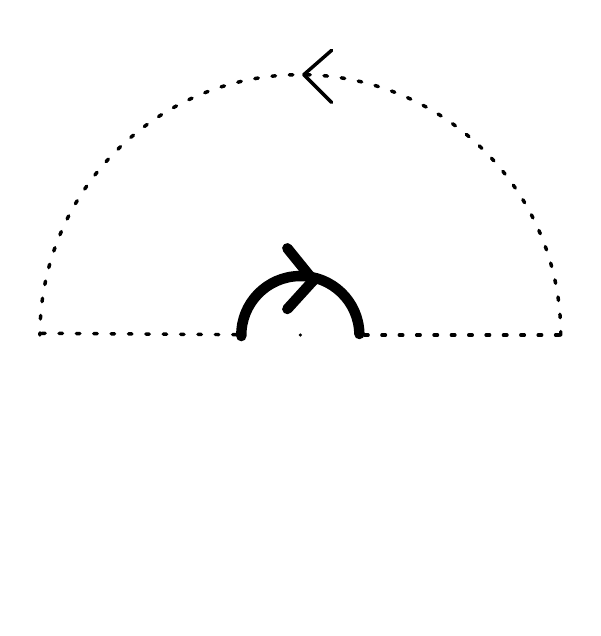}}{} ) dt \left( \frac{e^{2i x t}}{ 2\sinh bt \sinh b^{-1} t}- \frac{1}{2t^2} \right)  \,,
\ea
where we have assumed $\re(x)>0$ in the last line (else, we close the contour to the bottom).
The poles of the integrand lie at $t = i \pi m /b$ and $t= i \pi n b$, $m,n \in \IZ$, with 
\ba
\res_{t= i \pi m/b} \left( \frac{e^{2i x t}}{ 2\sinh bt \sinh b^{-1} t}- \frac{1}{2t^2} \right) &=& (-1)^m \frac{e^{-2  \pi xm/b}}{2 b \sinh \frac{i \pi m}{b^2}} \,, \\
\res_{t= i \pi n b} \left( \frac{e^{2i x t}}{ 2\sinh bt \sinh b^{-1} t}- \frac{1}{2t^2} \right) &=& (-1)^n \frac{b\,e^{-2  \pi x n b}}{2 \sinh i \pi n b^2}  \,,
\ea
for $m,n \neq 0$, and
\be
\res_{t= 0} \left( \frac{e^{2i x t}}{ 2\sinh bt \sinh b^{-1} t}- \frac{1}{2t^2} \right) = ix  \,.
\ee
Thus,
\be
\frac{d}{dx}\log s_b (x) = \frac{\pi}{i} \left( -x + \frac{1}{b} \sum_{m=1}^\infty (-1)^m \frac{e^{- 2 \pi x m/b}}{\sin \frac{ \pi m}{b^2}} + b \sum_{n=1}^\infty (-1)^n \frac{e^{-2 \pi x n b}}{\sin \pi n b^2} \right) \,.
\ee
To take the semi-classical limit, we drop the first sum, and approximate the sine-function in the second,
\ba
\frac{d}{dx}\log s_b (x) & =&  \pi i x - i \pi b \sum_{n=1}^\infty (-1)^n \frac{e^{- 2 \pi x n b}}{\pi n b^2 }   + O(b)\\
&=& \frac{i}{b} \log 2 \cosh \pi b x +  O(b) \,.
\ea
Integrating and imposing the boundary condition $s_b(0)=1$ then yields
\be
\log s_b(x)  = \frac{i}{b} \int_0^x dx' \log 2 \cosh \pi b x' + O(b)  \,.
\ee
To obtain the $\re(x) \rightarrow \pm \infty$ behavior of $s_b$, we can relate the right hand side of this approximation to the Lobachevsky function. Defining $S_b(x) = s_b(y)$ with $x = i y + \frac{Q}{2}$, we obtain
\be \label{sbint}
\log S_b(x) \approx \frac{1}{b} \int_{Q/2}^x dx' \log 2 \sin \pi b x'  \approx \frac{1}{\pi b^2} \int_{\frac{\pi}{2}}^{\pi b x} dy \log 2 \sin y\,.
\ee
The integral on the right hand side can be related to the dilogarithm function \cite{MR2249478}. This has the following integral definition:
\be
\Li_2(z) = \int_0^z \log(1-w) \frac{dw}{w}
\ee
for $|z| \le 1$. We choose the branchcut of the logarithm such that the function is analytic away from the semi-axis $[1, \infty)$. Note that for $w = e^{2i \xi}$, $-\frac{\pi}{2} < \re(\xi) < \frac{\pi}{2}$,
\be
\log(1-w) \frac{dw}{w} = \log \left(e^{i \xi}  (e^{-i \xi} - e^{i \xi}) \right) 2i \, d\xi = \left( i \xi + \log (-i) + \log 2\sin \xi \right)  2i \,d \xi  \,.
\ee
Choose the integration path depicted in the figure. 
\begin{figure}
\centering
\includegraphics[height=5cm]{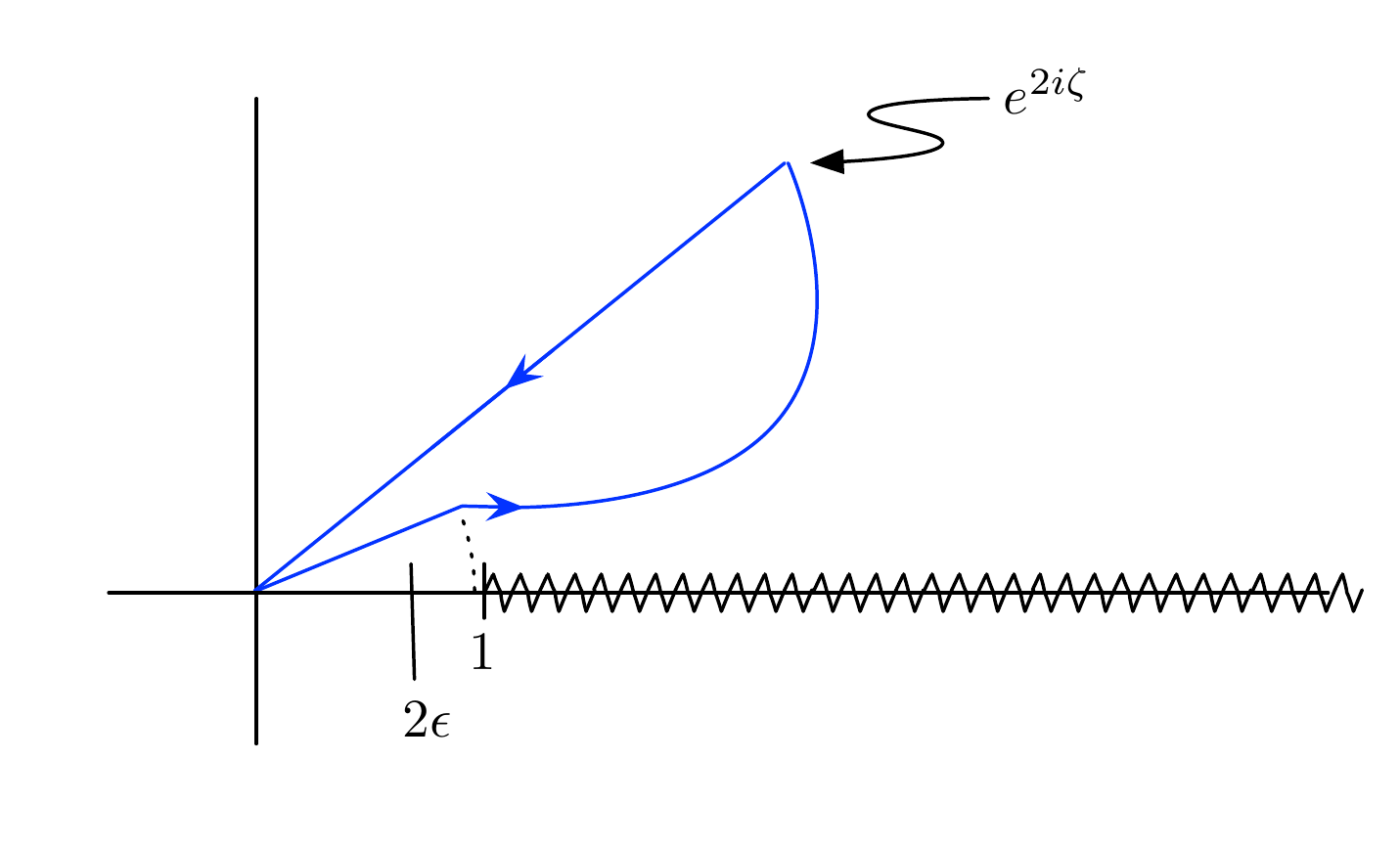}
\caption{Integration contour.}
\end{figure}
Then for $e^{2i \zeta} \not\in [1, \infty)$,
\ba
\Li_2(e^{2i \zeta}) - \Li_2 (e^{2i \epsilon}) &=& - \int_{\epsilon}^\zeta \left( i \xi + \log (-i) + \log 2\sin \xi \right)  2i \,d \xi \\
&=& - 2i \int_\epsilon^\zeta \log 2 \sin \xi \,d\xi - \pi (\zeta - \epsilon) +  \zeta^2- \epsilon^2 \,.
\ea
Taking the $\epsilon \rightarrow 0$ limit yields
\be
\int_0^\zeta \log 2 \sin \xi \,d\xi = \frac{i}{2} \left( \Li_2(e^{2i \zeta}) - \frac{\pi^2}{6} + \zeta ( \pi - \zeta) \right)  \,,
\ee
where we have used $\Li_2(1) = \frac{\pi^2}{6}$. We are interested in the $\im \,\zeta \rightarrow \pm \infty$ limits of this expression. The $\im \,\zeta \rightarrow +\infty$ limit follows immediately from $\Li_2(0) = 0$,
\be
\int_0^\zeta \log 2 \sin \xi \,d\xi \,\xrightarrow[\im \,\zeta \rightarrow + \infty]{} \,- \frac{i}{2} \left (\frac{\pi^2}{6} - \zeta ( \pi - \zeta) \right) \,.
\ee
 For the limit $\im \,\zeta \rightarrow \, - \infty$, note that
\be
\Li_2(z) + \Li_2(1/z) = - \frac{\pi^2}{6} - \frac{1}{2} \log^2(-z)  \,,
\ee
hence
\be
\lim_{\im \zeta \rightarrow - \infty}\Li_2 (e^{2i \zeta}) = -\frac{ \pi^2}{6} - \frac{1}{2} \log^2 e^{- i\pi + 2i \zeta} = \frac{ 2\pi^2}{6} - 2\zeta (\pi - \zeta)  \,.
\ee
To arrive at this result, remember that the branchcut of $\log z$ is chosen along the negative real axis, and that $-\frac{\pi}{2} < \re(\xi) < \frac{\pi}{2}$. Thus,
\be
\int_0^\zeta \log 2 \sin \xi \,d\xi \,\xrightarrow[\im \,\zeta \rightarrow - \infty]{} \, \frac{i}{2} \left( \frac{\pi^2}{6} - \zeta ( \pi - \zeta) \right) \,.
\ee
Substituting this result into (\ref{sbint}) and using
\be
\int_0^{\frac{\pi}{2}} dy \log 2 \sin y = 0
\ee
yields
\be
\log S_b(x) \xrightarrow[\im \, x \rightarrow \pm \infty]{b  \rightarrow 0} \mp\frac{\pi i}{2}  \left( x(x - \frac{1}{b}) + \frac{1}{6 b^2} \right)\,,
\ee
and thus
\be
\log s_b(x) \xrightarrow[\re \, x \rightarrow \pm \infty]{b  \rightarrow 0} \pm \frac{\pi i}{2} \left( x^2 + \frac{1}{12b^2}  \right)\,.
\ee

\bibliography{cft}
\bibliographystyle{utcaps}

\end{document}